\documentclass[twocolumn,secnumarabic,amssymb, nobibnotes,showpacs, aps, pra,10pt]{revtex4-1}
\usepackage{graphicx, epsfig,wrapfig,color}

\def\##1{{\mathbf{#1}}}
\def\=#1{\underline{\underline #1}}

\begin{document}

\title{Spontaneous decay of an emitter's excited state near a finite-length metallic carbon nanotube }%

\author{A. M. Nemilentsau, G. Ya. Slepyan, and S. A. Maksimenko}%
\address{Institute for Nuclear Problems, Belarus State University, Bobruiskaya 11, 220030 Minsk, Belarus}

\author{A. Lakhtakia}
\address{Department of Engineering Science and Mechanics, Pennsylvania State University, University Park, PA 16802, USA}

\author{S. V. Rotkin}
\address{ Department of Physics, and Center for Advanced Materials and Nanotechnology, Lehigh University, 16 Memorial Dr. E., Bethlehem, PA 18015, USA}

\begin{abstract}
The  spontaneous decay of an excited state of an emitter placed in the vicinity of a metallic single-wall carbon
nanotube (SWNT) was examined theoretically. The emitter-SWNT coupling strongly depends on
the position of the emitter relative to the SWNT, the length of the SWNT, the dipole transition frequency and the orientation of the emitter.
In the high-frequency regime,   dips in the spectrum of the spontaneous decay rate  exist
at the resonance frequencies in the spectrum of  the SWNT conductivity.   In the intermediate-frequency
regime, the SWNT conductivity is very low, and the spontaneous decay rate is practically
unaffected by the SWNT.
In the low-frequency regime,
the spectrum of the spontaneous decay rate contains resonances at the antennas resonance frequencies for
surface-wave propagation in the SWNT. Enhancement of both the {total} and radiative {spontaneous} decay rates by
several orders in magnitude is predicted at these resonance frequencies.
The strong emitter-field coupling is achieved, in spite of the low
Q factor of the antenna resonances, due to the very high magnitude of the electromagnetic field in the near-field zone. {The vacuum Rabi oscillations of the population of the excited emitter state are exhibited when the emitter is coupled to an antenna resonance of the SWNT}.
\end{abstract}


\date{Today}%
\maketitle

\section{Introduction}
The process of the spontaneous decay of an excited state of an emitter (atom, molecule, quantum dot, etc.)
is strongly affected by the surroundings the emitter placed in. In fact, any non-homogeneity in the medium surrounding an emitter---including the very simple case of an interface of two different
materials---will result in modification of the decay time (as well as the
angular distribution of the radiated power) due to the interaction with
the surface/interface modes \cite{purcell,shulga}.

The physical scenario becomes especially interesting
when the electromagnetic field (EMF)
modes are quantized in a resonator, an open resonator or another
photonic structure. Certain types of nanostructured resonators  have a strongly
inhomogeneous spatial or spectral distribution of the photonic density
of states (DOS). The dynamics of an emitter coupled to such a resonator
becomes non-Markovian \cite{Dalton,Walther,Kavokin,Petrov,Lambropoulos,Mogilevtsev,Vogel}.
Two relevant types of resonators or structured reservoirs are being studied these days:
microcavities \cite{Dalton,Walther,Kavokin} and photonic crystals
\cite{Petrov,Lambropoulos,Mogilevtsev}. Among notable phenomena
in the resonators are: (i) vacuum Rabi oscillations in the spontaneous decay of an atom coupled to a mode of a high-Q microcavity
\cite{Walther,Kavokin}, and (ii) complete freezing of the spontaneous decay process of an emitter with its transition frequency
inside the band gap of a photonic crystal \cite{Lambropoulos,Mogilevtsev}.

Antenna-like structures furnish another example of an open
resonator with a specific photonic DOS. One can consider
nanoantennas as structured photonic reservoirs of a new type, which
should act similarly to microcavities and photonic crystals.
Nanoantennas are also appealing as  new tools for controlling
emitter dynamics.

Nanoantennas of a particularly interesting type comprise nanoparticles made of plasmonic metals. For example, substantial enhancement of the fluorescence signal
of an emitter placed in the vicinity of such nanoparticles
\cite{Anger_2006,Bharadwaj_2007,Kuhn,Eghlidi,Mohammadi} or
nanoapertures \cite{Gerard} has been experimentally demonstrated.
The spontaneous decay of quantum dots has been suggested
\cite{Lukin_2006,Lukin_2007} to be an efficient way to generate
optical plasmons in metallic nanowires to which the quantum dots are
coupled. The crucial influence of antenna resonances in a gold
nanodisk on the decay rate of a nearby quantum emitter   has been
theoretically predicted \cite{Greffet}.

A single-wall carbon nanotube (SWNT) can function  as a
nanoantenna
\cite{Burke,Hanson_2005,Hao,Slepyan2006,Miano,Shuba_2007,Shuba_2009}. The cross-sectional diameter
of a SWNT is  of the order of a few nanometers while its length
may  be as large as  several centimeters. The structural symmetry of
the SWNT is denoted by a pair of indices $(m,n)$.
A SWNT may exhibit either
metallic/quasi-metallic properties (for $m-n = 3 q$, where $q$ is an
integer) or  semiconducting behavior ($m-n \neq 3 q$)
\cite{Charlier}.

The propagation of the surface electromagnetic waves \cite{Slepyan_99} along
the surface of a SWNT determines its scattering and radiation
characteristics \cite{Hanson_2005,Slepyan2006}. The antenna
resonances of the surface waves in a SWNT are the origin of the
pronounced peaks in the spectrum of its photonic DOS
\cite{Nemilentsau2009} and, correspondingly, in the spectrum of
thermal radiation if the SWNT is metallic \cite{Nemilentsau2007}. These
 resonances can be also seen in the polarizability tensor
and in the input impedance of a metallic SWNT. The resonance frequencies lie in the
terahertz/infrared regime,
depending on the SWNT length  \cite{Hanson_2005,Slepyan2006}.
These antenna resonances were shown to strongly affect the
electromagnetic coupling between a SWNT and a plasmonic
nanosphere \cite{Hanson_2007} or a dipole \cite{Nemilentsau2010}.
Their effect on the spontaneous decay of a nearby emitter is reported
here.

A strong enhancement of the spontaneous decay rate of an
excited atomic state was theoretically predicted  for an atom placed
in close proximity of the surface of an infinitely long SWNT
\cite{Bondarev_02,Bondarev_04}, although antenna effects were
neglected in these studies. We focus here on the combination
of the antenna and near-field effects, and their influence on the
process of the spontaneous decay of an emitter coupled to a SWNT
nanoantenna. Besides a fundamental theoretical interest, practical
applications can be foreseen, for example, in the fluorescence
microscopy with SWNT-based probes \cite{Hillenbrand,Mu}.
Moreover, the integration of SWNTs as nanoantennas with the
nanoscale luminescent materials such as quantum dots
\cite{Shi,Zhou,Wang} allows multifunctional nanostructures. Not only may these nanostructures
exhibit biocompatibility and fluorescence, but also other
characteristics useful for cancer diagnostics and therapeutics, as well as for drug
storage and delivery.

In Sec.
\ref{Sec:WeakCoupling} we consider the decay dynamics of the
emitter   in the weak-coupling regime. The characteristics of the
strong-coupling regime are discussed in Sec. \ref{Sec:Strong}
followed by conclusions in Sec. \ref{Sec:Conclusion}. Gaussian units
are used throughout, an $\exp(-i\omega t)$ dependence on time is
implicit, $\mathbf{e}_{x,y,z}$ are the unit Cartesian vectors, and all
tensors are of the second order.

\section{Weak-coupling regime: spontaneous emission}
\label{Sec:WeakCoupling}
\begin{figure}[ht!]
\begin{center}
\includegraphics[width=8.5cm]{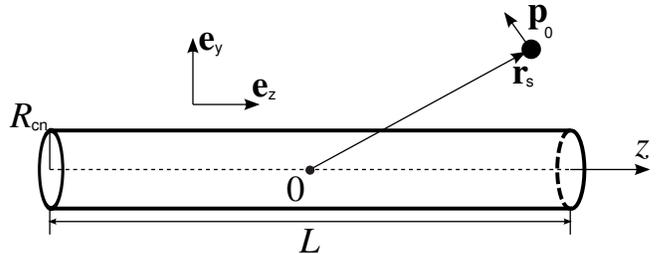}\\[2mm]
\small{} \caption{Schematic  representation of the emitter of electric
dipole moment ${\mathbf p}_0$ located at ${\mathbf r}_s$ in the
vicinity of a SWNT of length $L$ and cross-sectional radius
$R_{cn}$. The origin $0$ of the coordinate system lies at the centroid
of the SWNT and the $z$ axis is parallel to the axis of the SWNT. }
\label{Fig:CNT}
\end{center}
\end{figure}

Let us consider the spontaneous decay of
an excited state of an emitter  placed in the vicinity of
a SWNT of finite length $L$, as shown in Fig. \ref{Fig:CNT}. The emitter, located at
$\mathbf{r}_s$, is modeled by a two-level system with transition
frequency $\omega_e$ and dynamic electric dipole moment
$\mathbf{p}_0$. In the weak-coupling regime, the decay process is
Markovian and the evolution of the excited
state's occupation probability $|C_u(t)|^2$ is described by the exponential law
\cite{Vogel,Bondarev_02,Bondarev_04}:
\begin{equation} \label{Eq:Decay_MarKovian}
|C_{u}(t)|^2 = e^{-\Gamma_{cn}(\mathbf{r}_s,\omega_e) t},
\end{equation}
where
\begin{eqnarray}
&& \Gamma_{cn}(\mathbf{r}_s,\omega_e)  =   \frac{2 \omega_e^2}{\hbar c^2} \, \mathbf{p}_0  \cdot\mathrm{Im} \left[\underline{\underline{G}}
(\mathbf{r}_s,\mathbf{r}_s,\omega_e) \right]\cdot\mathbf{p}_0   \nonumber \\
\label{Eq:Decay_rate}
&&\quad  =  \Gamma_0 \left\{1 + \frac{3 c}{2 \omega_e}\, \hat\mathbf{p}_0 \cdot \mathrm{Im}\left[
\underline{\underline{G}}^{(sc)} (\mathbf{r}_s,\mathbf{r}_s,\omega_e)\right] \cdot \hat\mathbf{p}_0\right\}
\end{eqnarray}
is the spontaneous decay rate, $\Gamma_0 = 4 \omega_e^3
|\mathbf{p}_0|^2 /3\hbar c^3$ is the free-space decay rate,
$\hat\mathbf{p}_0$ is the unit vector along the direction of polarization of the  electric dipole, $c$ is the speed of light
in vacuum, $\hbar$ is the reduced Plank's constant and the Green tensor
$\underline{\underline{G}}(\mathbf{r},\mathbf{r}';\omega)$
is the solution
of the Fredholm integral equation (\ref{integral_equation}). {Equation~(\ref{Eq:Decay_rate}) defines the total spontaneous decay rate of
the emitter and includes both the radiative and the non-radiative decay rates.}

\begin{figure}[ht!]
\begin{center}
\includegraphics[width=8.5cm]{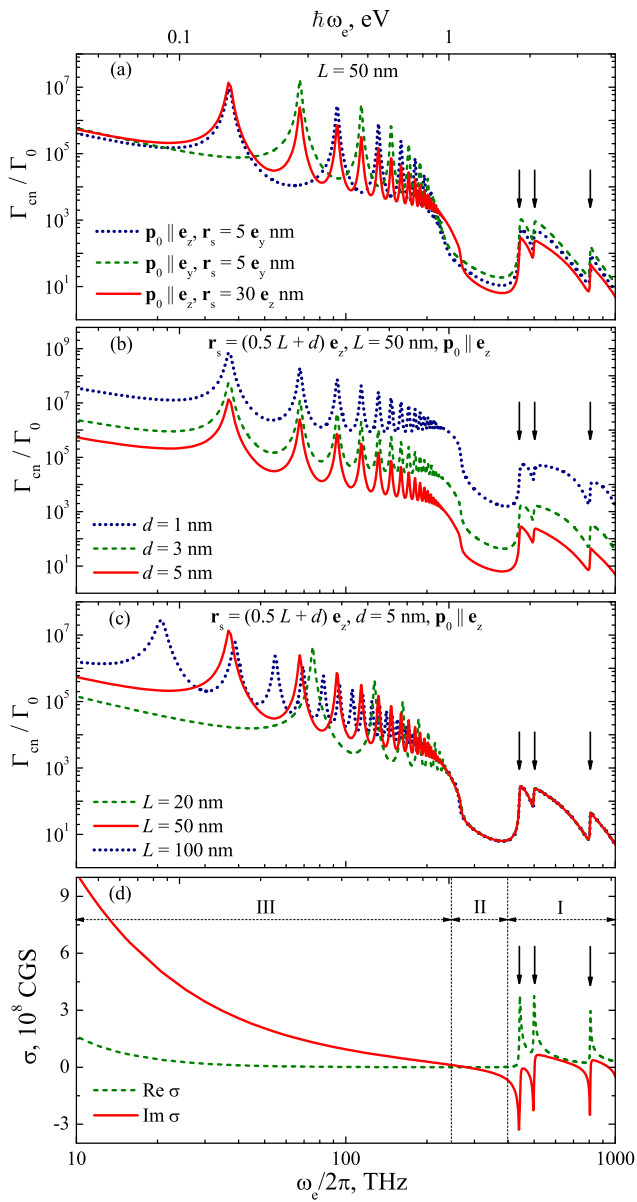}\\[2mm]
\small{}
\caption{Color online. (a)-(c) Normalized spontaneous decay rate $\Gamma_{cn}/\Gamma_0$ of an emitter coupled to a metallic $(15,0)$
SWNT, as in Fig. \ref{Fig:CNT}, for (a) different positions $\mathbf{r}_s$ and orientations $\mathbf{p}_0/\vert\mathbf{p}_0\vert$ of the emitter,
assuming that the distance $d$ between the emitter and the point on SWNT
surface closest to the emitter remains the same; (b) different distances $d$; (c) different lengths $L$ of the SWNT. Equation (\ref{Eq:Decay_rate}) was used for
calculating the decay rate.
(d) Linear conductivity of the $(15,0)$ metallic SWNT calculated using
Eq. (36) in Ref.~\onlinecite{Slepyan_99}. The electronic relaxation time is assumed to be equal to $10^{-13}$~s.
}
\label{Fig:Decay_rate_metallic}
\end{center}
\end{figure}

The spectra of the  {normalized spontaneous decay rate $\Gamma_{cn}/\Gamma_0$} of an excited state of the emitter coupled to a metallic $(15,0)$ SWNT are presented in Fig. \ref{Fig:Decay_rate_metallic} for different locations and orientations of the emitter (Fig. \ref{Fig:Decay_rate_metallic}a), different distances between the emitter and the SWNT (Fig. \ref{Fig:Decay_rate_metallic}b), and different SWNT lengths (Fig. \ref{Fig:Decay_rate_metallic}c).
The spectra of the real and imaginary parts of the conductivity of the SWNT are presented in Fig. \ref{Fig:Decay_rate_metallic}d. We can identify three regimes
in the conductivity spectrum:
\begin{itemize}
\item[(I)] the high-frequency regime, $\omega_e/(2\pi)> 400$ THz, wherein the optical interband transitions dominate the SWNT conductivity;
\item[(II)] the intermediate-frequency regime,
$250 < \omega_e/(2\pi) < 400$ THz; and
\item[(III)] the low-frequency regime, $\omega_e/(2\pi) < 250$ THz, wherein the conductivity follows the Drude model.
\end{itemize}

Let us consider  {first} the high-frequency regime I in Fig.~\ref{Fig:Decay_rate_metallic}.
The spontaneous decay rate demonstrates a significant dependence on the shortest
distance $d$ between the emitter and the closest point on the SWNT surface  (Fig.~\ref{Fig:Decay_rate_metallic}b),
thereby revealing the crucial influence of the near fields, as discussed later with reference to   Fig. \ref{Fig:Figure3}b.

The conductivity spectrum  {shows fixed resonances}, identified
by arrows in Fig. \ref{Fig:Decay_rate_metallic}, that originate from
the electron interband transitions between the Van Hove singularities in the electronic DOS.
The spectra of the spontaneous decay rate of the emitter show
dips at these frequencies.  {The frequencies of dips depend on the SWNT index
$(m,n)$, but not on the length of the SWNT.} Furthermore,
the dip  { frequencies} do not show any dependence on either the position
or the orientation of the emitter. The
magnitude of the spontaneous decay rate is  affected by the
orientation of the emitter (Fig.~\ref{Fig:Decay_rate_metallic}a) and
is independent of  $L$ (Fig.~\ref{Fig:Decay_rate_metallic}c). Thus
we conclude that these near-field modes are not related to plasmonic
resonances whose frequencies would scale with   $L$.

In the intermediate-frequency regime II the conductivity of the SWNT
is small and its antenna response is weak. Hence, the
spontaneous decay rate is lower than in the regimes I and III.
Actually, the spontaneous decay rate is almost the same as the
free-space decay rate  (i.e., in the
absence of the SWNT) when $d>5$~nm. Our
predictions for the decay dynamics in the frequency regimes I and II
coincide
with the results presented in Refs.~\onlinecite{Bondarev_02} and \onlinecite{Bondarev_04} that were obtained for infinitely long SWNTs.

Most interesting is the low-frequency regime III, where the
theoretical predictions made assuming the SWNT is infinitely long
\cite{Bondarev_02,Bondarev_04} cannot explain the results.
Particularly, Figs. \ref{Fig:Decay_rate_metallic}a-c indicate the
presence of several resonances instead of the monotonic decrease
of the spontaneous decay rate with the decrease of the transition
frequency of the emitter, as  predicted in
Refs.~\onlinecite{Bondarev_02} and \onlinecite{Bondarev_04}. The
frequencies of these resonances coincide with the frequencies of
the antenna resonances (arising from surface-wave propagation in
the SWNT) defined by the  {space-quantization} condition {for
plasmons} \cite{Slepyan2006}:
\begin{equation} \label{Eq:resonance_freq}
h L \approx \pi s,
\end{equation}
where $h$ is the guide wavenumber that is calculated using Eq.~(58)
of Ref.~\onlinecite{Slepyan_99} and $s$ is an integer. Furthermore, the location
and the orientation of the emitter as well as the length of the SWNT
strongly affect the spontaneous decay rate. In particular,  {only
one type of resonances, with $s$ being either odd for all resonances or even for all resonances, is present in
the spectrum of the spontaneous decay rate if the emitter is located
in the middle of the tube, {\em i.e.,}} equidistant from both SWNT
edges (Fig. \ref{Fig:Decay_rate_metallic}a, $\mathbf{r}_s = 5
\mathbf{e}_y$ nm). It  depends on the orientation of the emitter
with respect to the SWNT axis, which reflects odd or even symmetry
of the dipole potential of the emitter. However, when the emitter is
located {on the SWNT axis, near one of the two edges of the SWNT
and also oriented along the  axis, both odd-$s$ and even-$s$ resonances are
present. The symmetry in this case is full axial, although the mirror
symmetry with respect the SWNT center is broken.

\begin{figure}[ht!]
\begin{center}
\includegraphics[width=8.5cm]{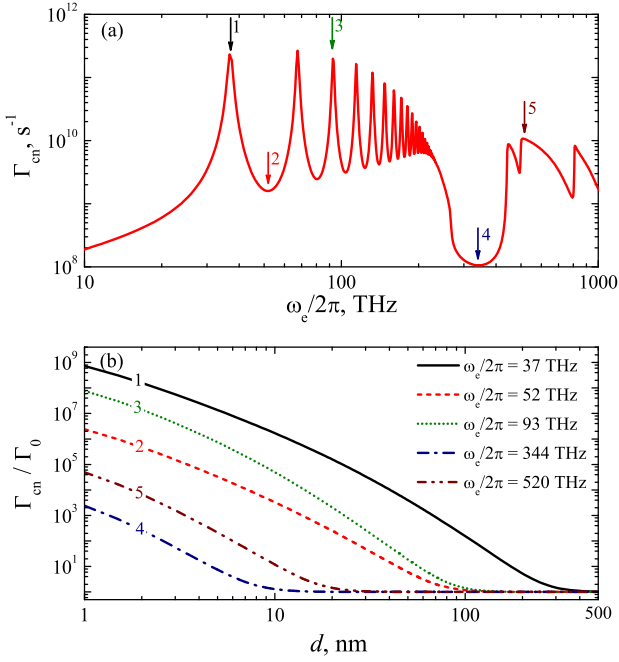}\\[2mm]
\small{} \caption{Color online. (a) Spontaneous decay rate $\Gamma_{cn}$
of an emitter with electric dipole moment  $\mathbf{p}_0 = 1\mathbf{e}_z$ debye placed in the vicinity of an edge ($\mathbf{r}_s = 30 \mathbf{e}_z$ nm) of a $(15,0)$ metallic SWNT of length $L=50$ nm. The corresponding spectrum of the normalized decay rate $\Gamma_{cn}/\Gamma_0$   is presented as
the solid line in Fig. \ref{Fig:Decay_rate_metallic}b. (b) Dependence of  $\Gamma_{cn}/\Gamma_0$ on the separation $d$ between the SWNT edge and the emitter for different emitter frequencies indicated by arrows in Fig. \ref{Fig:Figure3}a, when $\mathbf{r}_s = (L/2+d) \mathbf{e}_z$ nm, $L = 50$ nm, and $\mathbf{p}_0 \parallel \mathbf{e}_z$.
}
\label{Fig:Figure3}
\end{center}
\end{figure}

 The spontaneous
decay of the emitter strongly depends on the length of the SWNT.
A decrease of $L$ leads to the blue shift of the resonance
frequencies (Fig. \ref{Fig:Decay_rate_metallic}a), according to the space-quantization
condition (\ref{Eq:resonance_freq}) for plasmons.
Nevertheless, even for a short SWNT ($L\sim20$~nm), at least a few
resonances appear before the frequency regime II begins, where the
resonances are not observable due to the strong attenuation of
surface waves. Similar influence of the antenna resonances on the
decay rate has been theoretically demonstrated for the case of an
emitter coupled to a plasmonic (silver) nanodisk in Ref.~\onlinecite{Greffet}.

Whereas Fig.~\ref{Fig:Decay_rate_metallic} contains spectra of the normalized spontaneous decay rate $\Gamma_{cn}/\Gamma_0$,
the spectra of the actual  spontaneous decay rate $\Gamma_{cn}$  are interesting
in their own right. Let us compare the spectrum of $\Gamma_{cn}$ in Fig.~\ref{Fig:Decay_rate_metallic} with the spectrum
of $\Gamma_{cn}/\Gamma_0$ depicted as a solid line
in Fig. \ref{Fig:Decay_rate_metallic}b, both plots having been drawn for the same emitter-SWNT configuration.
The off-resonance values of   $\Gamma_{cn}$ increase, but those of $\Gamma_{cn}/\Gamma_0$ decrease,
as  $\omega_e$ increases. Because $\Gamma_0\propto\omega_e^3$ by definition, the off-resonance values of $\Gamma_{cn}
\sim\omega_e^\alpha$, $\alpha\in(0,3)$.

The dependence of $\Gamma_{cn}/\Gamma_0$ on the separation $d$ between the emitter and the nearest point on the SWNT
surface is presented in Fig. \ref{Fig:Figure3}b for five  different values of  $\omega_e$ identified in Fig. \ref{Fig:Figure3}a.
The enhancement indicated by $\Gamma_{cn}/\Gamma_0>1$  fades rapidly as $d$ increases, which shows the crucial
effect of the high magnitude of the electric field in the near-field zone.  Let $d_0$ denote the value of $d$ for which
$\Gamma_{cn}$ becomes equal to $\Gamma_0$ as $d$ increases.
Then, at the frequency of the first antenna resonance (curve
labeled 1 in Fig. \ref{Fig:Figure3}b),   $d_0 \approx 0.1\pi{c}/\omega_e$; furthermore,   $d_0$   decreases non-monotonically as $\omega_e$ increases. The values of $d_0$ at the antenna resonance frequencies of the SWNT are higher than the ones at
the off-resonance frequencies. We also observe strong decrease of $d_0$ in the intermediate-frequency regime II.

\begin{figure}[ht!]
\begin{center}
\includegraphics[width=8.5cm]{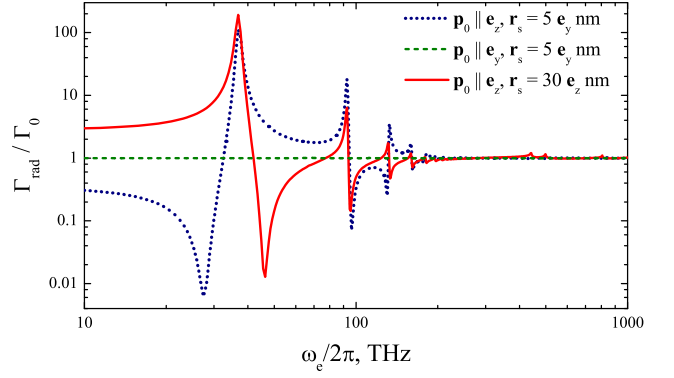}\\[2mm]
\small{} \caption{Normalized spontaneous radiative decay rate $\Gamma_{\mathrm{rad}}/\Gamma_0$ of the emitter coupled to
a  $(15,0)$ metallic SWNT for the same  configurations as in Fig. \ref{Fig:Decay_rate_metallic}a.  }
\label{Fig:Figure4}
\end{center}
\end{figure}

 The decay rate $\Gamma_{cn}$ defined by Eq. (\ref{Eq:Decay_rate}) and depicted in Figs. \ref{Fig:Decay_rate_metallic}
and \ref{Fig:Figure3} comprises a radiative part $\Gamma_{\mathrm{rad}}$ and a non-radiative part $\Gamma_{\mathrm{nr}}$
as follows:
\begin{equation}
\Gamma_{cn}  = \Gamma_{\mathrm{rad}}+\Gamma_{\mathrm{nr}}.
\end{equation}
The spontaneous radiative decay rate is given by \cite{Lukin_2006,Novotny}:
\begin{equation} \label{Eq:Radiative_decay}
\Gamma_{\mathrm{rad}} = \eta \Gamma_{cn},
\end{equation}
where $\eta$ is the radiation efficiency of the emitter in the presence of SWNT, as defined by Eq. (\ref{Eq:radiation_efficiency}).

The spectra of $\Gamma_{\mathrm{rad}}/\Gamma_0$ are presented in Fig. \ref{Fig:Figure4} for the same  configurations as in Fig. \ref{Fig:Decay_rate_metallic}a. Clearly, $\Gamma_{\mathrm{rad}}$ is enhanced much less than $\Gamma_{cn}$ is.
In fact, in the frequency regimes I and II we do not observe any enhancement of $\Gamma_{\mathrm{rad}}$ due to the presence
of the SWNT.

In the frequency region III, resonances arise in the  spectrum of $\Gamma_{\mathrm{rad}}$ provided the emitter
is polarized parallel to the SWNT axis ($\mathbf{p}_0 || \mathbf{e}_z$), the resonance frequencies being equal to the frequencies of antenna resonances defined by Eq. (\ref{Eq:resonance_freq}).
The enhancement of the spontaneous radiative decay rate can be as high as 100 at the first resonance and falls rapidly as the resonance number $s$ increases. When the emitter is placed near an edge of the SWNT, only odd-$s$ resonances arise in the spectrum of $\Gamma_{\mathrm{rad}}$,
(solid line in Fig. \ref{Fig:Figure4}), though both odd-$s$ and even-$s$ resonances are present in the spectrum of $\Gamma_{cn}$
(solid line in Fig. \ref{Fig:Decay_rate_metallic}a). The reason of the even-$s$ resonances disappearance is the asymmetric distribution of the electric current induced in the SWNT by the emitter relative to the centroid of the SWNT that leads to the strong attenuation of the electromagnetic field radiated by the SWNT in the far-field zone. This is also the reason for the absence of resonances in the  spectrum of
$\Gamma_{\mathrm{rad}}/\Gamma_0$ when the emitter is polarized normal to the SWNT axis (dashed line in Fig. \ref{Fig:Figure4}).

\begin{figure}[ht!]
\begin{center}
\includegraphics[width=8.5cm]{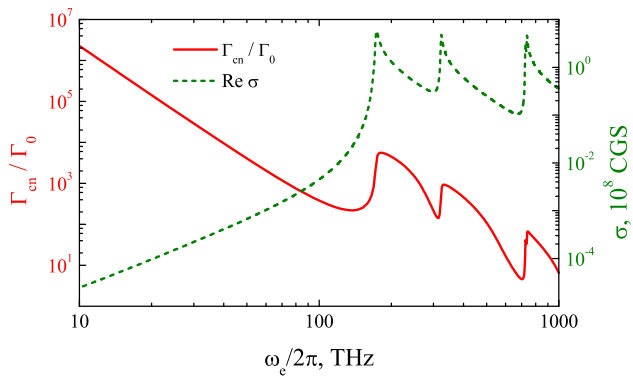}\\[2mm]
\small{}
\caption{Color online. Spontaneous decay rate of an emitter placed in the vicinity of a semiconducting (14,0) SWNT of length $L=50$ nm; $\mathbf{p}_0 || \mathbf{e}_z$ and
$\mathbf{r}_s = 30 \mathbf{e}_z$ nm. }
\label{Fig:Decay_rate_semiconductor}
\end{center}
\end{figure}

In contrast to metallic SWNTs, the room-temperature conductivity of
{an intrinsic (undoped)} semiconducting SWNT is low due to the
low density of electrons in the conduction band. This leads to a strong
attenuation of surface waves and, correspondingly, to the absence of
resonances in the spectrum of the spontaneous decay rate {as
presented in Fig.~\ref{Fig:Decay_rate_semiconductor}. In the frequency regime I
the contribution of the interband electron transitions dominates the
SWNT conductivity. Then the spectrum of the spontaneous decay rate
has similar features for both metallic and semiconducting SWNTs, \textit{i.e.},
the dips are present at the resonance frequencies of the
conductivity.

\section{Strong-coupling regime: vacuum Rabi oscillations}
\label{Sec:Strong}

By virtue of  Eq. (\ref{Eq:Decay_rate}) and the spectra of
$\Gamma_{cn}/\Gamma_0$  presented in Fig. \ref{Fig:Decay_rate_metallic},
we conclude that, in the presence of the (15,0) metallic SWNT, the
spectra of the imaginary part of the diagonal elements of
$\underline{\underline{G}} (\mathbf{r}_s,\mathbf{r}_s,\omega)$
must have pronounced resonances. So must the photonic DOS, as it is
proportional to the trace of
$\mathrm{Im}\underline{\underline{G}}\left[(\mathbf{r}_s,\mathbf{r}_s,\omega)\right]$.
Accordingly, a metallic SWNT can be considered as  {an open
resonator (a structured reservoir) with a specific spectrum} of
quantized modes of the EMF.
We {suggest} that the {physics} of an emitter coupled to {these modes should be similar}
to  {the case of} an emitter coupled to a microcavity \cite{Dalton} or a photonic crystal
\cite{Lambropoulos}: in particular, the Rabi oscillations of the population of the
emitter's excited state {are expected}.

When  the dipole transition frequency of the emitter,
$\omega_e/2\pi$, is close to the frequency of a resonance of the
photonic DOS, $\omega_{\nu}/2\pi$,  {a strong
coupling develops between the field and the emitter.} Assuming that
the resonance in the photonic DOS spectrum can be
approximated by a Lorentzian shape, we write:
\begin{equation} \label{Eq:Lorentz_fit}
\mathbf{p}_0  \cdot\mathrm{Im}
\left[\underline{\underline{G}}
(\mathbf{r},\mathbf{r},\omega)\right] \cdot\mathbf{p}_0 =  \frac{ \mathbf{p}_0  \cdot\mathrm{Im}\left[ \underline{\underline{G}}
(\mathbf{r},\mathbf{r},\omega_{\nu}) \right]\cdot\mathbf{p}_0 \,\gamma_{\nu}^2}{(\omega-\omega_{\nu})^2+\gamma_{\nu}^2}\,;
\end{equation}
furthermore,  {we assume that} the line width $\gamma_{\nu}$ is
sufficiently small {compared with the spectral separation between two
adjacent resonance lines \cite[Section 10.1.2]{Vogel}. {Then
the amplitude of the excited state is obtained as
\begin{equation} \label{Eq:Decay_strong}
C_u(t) = \frac{1}{\lambda_1-\lambda_2}\left(-\lambda_2 e^{\lambda_1 t} + \lambda_1 e^{\lambda_2 t}\right),
\end{equation}
where the {temporal decay constants are
\begin{equation}
\lambda_{1,2} = \frac{1}{2}(i \delta_{\nu} - \gamma_{\nu}) \pm \frac{1}{2} \sqrt{(i \delta_{\nu} - \gamma_{\nu})^2 -
2\Gamma(\mathbf{r}_s,\omega_e) \gamma_{\nu} }
\end{equation}
and  {the detuning is} $|\delta_{\nu}| = |\omega_e -
\omega_{\nu}| \ll \omega_{e,\nu}$. For $\gamma_{\nu} \gg
\Gamma(\mathbf{r}_s,\omega_e)$ we still get the exponential
decay of an excited state with decay rate
$\Gamma(\mathbf{r},\omega_e)$. However, for  $
2\Gamma(\mathbf{r}_s,\omega_e) > \gamma_{\nu}$ the
Markovian approximation fails and the temporal evolution of the occupation
probability amplitude of the excited state becomes
{non-monotonic}.

\begin{figure}[ht!]
\begin{center}
\includegraphics[width=8.5cm]{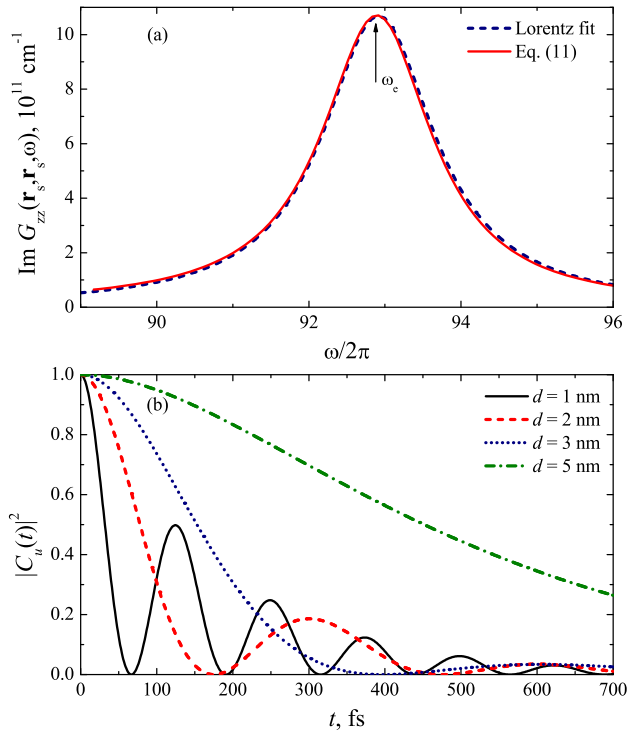}\\[2mm]
\small{}
\caption{Color online. (a) Solid line: $\#e_z\cdot\mathrm{Im} \left[\underline{\underline{G}}(\mathbf{r}_s,\mathbf{r}_s,\omega)\right]\cdot\#e_z$ calculated using
Eqs. (\ref{Eq:Greent_tensor_SWNT})--(\ref{integral_equation})
at $\mathbf{r}_s = 26 \mathbf{e}_z$ nm near an edge of a (15,0) SWNT of length $L = 50$ nm. Dashed line: Lorentzian fit  per
Eq. (\ref{Eq:Lorentz_fit}) with $\omega_{\nu} = 5.84 \times 10^{14}$ rad/s and $\gamma_{\nu} = 5.6 \times 10^{12}$ rad/s.
(b) The occupation probability $|C_u(t)|^2$ of an emitter excited state.
The angular frequency  $\omega_e$ of the dipole transition, designated by an arrow in Fig. \ref{Fig:Decay_Rabi}a, is taken to be
equal to $\omega_{\nu}$, i.e. $\delta_{\nu} = 0$. The transitional dipole moment of the emitter is   $\mathbf{p}_0 = 10 \,\mathbf{e}_z$ debye. }
\label{Fig:Decay_Rabi}
\end{center}
\end{figure}

Let us consider the spontaneous decay of an emitter  placed in the
vicinity of a (15,0) metallic SWNT of length $L = 50$ nm and assume
that the dipole transition frequency  $\omega_e$ is close to the
resonant frequency $\omega_{\nu}$.
Figure~\ref{Fig:Decay_rate_metallic} shows that the coupling of the
emitter to the SWNT strongly depends on both the location and the
orientation of the emitter in the low-frequency regime III. Suppose
that $\mathbf{p}_0 = 10\, \mathbf{e}_z$ debye so that the emitter is
oriented parallel to the SWNT axis, and that the emitter is placed
close to one edge of the SWNT such that $\mathbf{r}_s = (L/2 +
d)\,\mathbf{e}_z= 26 \,\mathbf{e}_z$ nm. The spectrum of
$\mathrm{Im}\left[\#e_z\cdot\=G(\mathbf{r}_s,\mathbf{r}_s,\omega)\cdot\#e_z\right]$
at $\mathbf{r}_s = 26 \,\mathbf{e}_z$ nm is presented in Fig.
\ref{Fig:Decay_Rabi}a  {near the resonance and is fitted by the
Lorentzian function (\ref{Eq:Lorentz_fit}) with the parameters:
$\omega_{\nu} = 5.84 \times 10^{14}$ rad/s and $\gamma_{\nu}
= 5.6 \times 10^{12}$ rad/s. Even though the Q factor of this
resonance is $\sim 100$, which is much lower than in microcavities
\cite{Kavokin}, we} can still achieve a strong emitter-field coupling
regime with an SWNT, due to the very high magnitude of the
EMF in the near-field zone.

The temporal evolution of the emitter excited state occupation probability
is presented in Fig. \ref{Fig:Decay_Rabi}b for different
emitter-SWNT separations $d$, assuming that the detuning
$\delta_{\nu}$ is null-valued. For small emitter-SWNT separation
distances ($d < 3$ nm), $C_u(t)$ provids an evidence of damped
Rabi oscilllations. The oscillation frequency decreases  as the
separation distance increases. Finally, for $d>5$~nm, the
spontaneous decay state becomes almost exponential. Thus, the
strong emitter-field coupling can be detected for an emitter in the
vicinity of a metallic SWNT acting {as a resonator, although} the
coupling strength   crucially depends on the emitter-SWNT separation.
The coupling strength can be increased by using shorter SWNTs, as
the blue-shift of the antenna resonance frequency  leads to an
increase in the Q factor.

\section{Discussion and conclusion}
\label{Sec:Conclusion}

Using the standard procedure of electromagnetic-field quantization in
an absorbing medium \cite{Vogel,Bondarev_02,Bondarev_04}, we
theoretically studied the process of spontaneous decay of an excited
state of an emitter placed in the vicinity of a metallic single-wall
carbon nanotube. We demonstrated that a metallic SWNT  can act
as  {an open resonator (a structured reservoir) with a special
density of quantized modes of the EMF in its vicinity}.
The emitter-SWNT coupling strongly depends on the position {and
the orientation} of the emitter relative to the SWNT, the length of the
SWNT, and the dipole transition frequency.

Three different frequency regimes were identified, based on the
characteristics  of the emitter-SWNT coupling. In the high-frequency
regime, where the dominant contribution to the SWNT conductivity is
due to the interband electron transitions, dips in the spectrum of the
spontaneous decay rate (Fig. \ref{Fig:Decay_rate_metallic}) exist at
the resonance frequencies {of the interband transition}. The
spontaneous decay rate of an excited state of the emitter is hardly
influenced by  the length of the SWNT {in this regime, although it does
depend on} the location and the orientation of the emitter. In the
intermediate-frequency regime, the SWNT conductivity is low, and
the spontaneous decay rate is practically unaffected by the SWNT.

In the low-frequency regime, the intraband motion of electrons
dominates the SWNT conductivity. The spectrum of the spontaneous
decay rate  {shows resonances coinciding with} the antenna
resonance frequencies for {electromagnetic waves (plasmons),
propagating} on the SWNT surface. Enhancement of the spontaneous decay
rate by at least 7 orders in magnitude is predicted at the resonance
frequencies (Fig. \ref{Fig:Decay_rate_metallic}).

 The contribution of the radiative decay to the spontaneous decay was estimated. The enhancement of the
spontaneous radiative decay rate by up to two orders of
magnitude was demonstrated for an emitter polarized along the SWNT axis. The spontaneous  radiative decay
rate was not enhanced when the emitter is
 polarized normal to the SWNT axis.

When the emitter is placed  several nanometers from the SWNT
surface and the dipole transition frequency of the emitter is in the
vicinity of an antenna resonance frequency of the SWNT,  the emitter
decay dynamics becomes non-Markovian. In particular, we clearly
{demonstrated
vacuum Rabi oscillations (Fig. \ref{Fig:Decay_Rabi}). The strong emitter-field coupling is achieved, in spite of the relatively
 low
Q factor of the antenna resonances, due to the very high magnitude of the electromagnetic field in the near-field zone.
Consequently, unlike a microcavity reservoir whose resonances have
high Q factors, a metallic SWNT nanoantenna should radiate quite
{readily}.

\acknowledgements
This research was partially supported by (i) the Belarus Republican Foundation for Fundamental Research (BRFFR) under projects F10R-004 and F10Mld-003; (ii)  EU FP7 under projects FP7-230778 TERACAN,  FP7-247007 CACOMEL and  FP7-266529 BY-NanoERA; and (iii) the Binder Endowment at the Pennsylvania State University.

\appendix
\section{Green tensor}

\label{Sec:theory_green}

The  Green tensor
$\underline{\underline{G}}(\mathbf{r},\mathbf{r}_s,\omega) $ in
Eq.~(\ref{Eq:Decay_rate}) is the solution of the differential
equation
\begin{equation} \label{Eq:Green_tensor_eq}
\left[\left(\nabla_{\mathbf{r}}\times\=I\right)\cdot\left(\nabla_{\mathbf{r}}\times\=I\right)-k^2\=I\right]\cdot\=G(\#r,\mathbf{r}_s,\omega)
=
4 \pi \underline{\underline{I}} \delta(\mathbf{r}-\mathbf{r}_s)
\end{equation}
which also satisfies  jump conditions appropriate for the EMF across the
SWNT surface as well as the Sommerfeld radiation conditions; here,
$\=I$ is the identity tensor,
$\delta(\cdot)$ is the Dirac delta function, and $k=\omega/c$ is free-space wavenumber.

Defining the vector field
$\#G^{(\beta)}(\#r,\#r_s,\omega)=\=G(\#r,\#r_s,\omega)\cdot\#e_\beta$,
$(\beta=x,y,z)$, we write:
\begin{equation} \label{Eq:Green_vector}
\left[\left(\nabla_{\mathbf{r}}\times\=I\right)\cdot\left(\nabla_{\mathbf{r}}\times\=I\right)-k^2\=I\right]\cdot\#G^{(\beta)}(\#r,\#r_s,\omega)
=
4 \pi \#e_\beta\delta(\mathbf{r}-\mathbf{r}_s)\,.
\end{equation}
Thus, we can formally consider each column of the Green tensor
$\=G(\#r,\#r_s,\omega)$ as a vector field
$\#G^{(\beta)}(\#r,\#r_s,\omega)$ induced at $\mathbf{r}$ by a
source current density located at $\mathbf{r}_s$ and polarized
parallel to the unit vector $\mathbf{e}_{\beta}$. Identifying the
electric field $\#E(\#r)$ as $\#G^{(\beta)}(\#r,\#r_s,\omega)$ and
the magnetic field $\#H(\#r)$ as
$(ik)^{-1}\nabla_{\#r}\times\#G^{(\beta)}(\#r,\#r_s,\omega)$,
we see that $\#G^{(\beta)}(\#r,\#r_s,\omega)$ satisfies two  jump
conditions across the SWNT surface \cite{Slepyan_99}; thus,
\begin{widetext}
\begin{eqnarray}
&& \lim_{\delta \rightarrow 0} \left\{ \mathbf{u}_{n}\times \nabla \times\left[\mathbf{G}^{(\beta)}(\mathbf{r}_{cn} + \delta \mathbf{u}_{n},\mathbf{r}_s,\omega)
 -\,\mathbf{G}^{(\beta)}(\mathbf{r}_{cn} - \delta \mathbf{u}_{n},\mathbf{r}_s,\omega) \right]\right\}  = \frac{4\pi ik^2}{\omega}
\mathbf{j}^{(\beta)}(z,\mathbf{r}_s), \label{Eq:boundary1}  \\
&&\lim_{\delta \rightarrow 0} \left\{ \mathbf{u}_{n}  \times \left[\mathbf{G}^{(\beta)}(\mathbf{r}_{cn} + \delta \mathbf{u}_{n},\mathbf{r}_s,\omega)  -\,
\mathbf{G}^{(\beta)}(\mathbf{r}_{cn} - \delta \mathbf{u}_{n},\mathbf{r}_s,\omega) \right]\right\}  = \#0, \label{Eq:boundary2}
\end{eqnarray}
where
\begin{equation} \label{Eq:Surface_vector}
\left.\begin{array}{l}
\mathbf{r}_{cn} = R_{cn} \#u_n+ z\, \mathbf{e}_z\\
\#u_n= \cos\phi \, \mathbf{e}_x+  \sin\phi\,\mathbf{e}_y
\end{array}\right\},
\end{equation}
the surface current density
$\mathbf{j}^{(\beta)}(z,\mathbf{r}_s,\omega) = \mathbf{e}_z \,
j^{(\beta)}(z,\mathbf{r}_s,\omega) =
\sigma_{zz}\,\mathbf{e}_z\mathbf{e}_z\cdot
\mathbf{G}^{(\beta)}(\mathbf{r}_{cn},\mathbf{r}_s,\omega)$ is
induced on the SWNT surface by the electric field
$\mathbf{G}^{(\beta)}(\mathbf{r}_{cn},\mathbf{r}_s)$, and
$\sigma_{zz}$ is the axial SWNT conductivity \cite{Slepyan_99}. We impose the restriction $ k R_{cn} \ll 2 \pi$ to ensure that $j^{(\beta)}(z,\mathbf{r}_s,\omega)$
is uniform along the SWNT circumference.
\end{widetext}

So, the calculation of the   Green tensor
$\underline{\underline{G}}(\mathbf{r},\mathbf{r}_s,\omega) $
requires us to solve three scattering problems. Each problem involves
the scattering  {by the SWNT of the field radiated by an electric
dipole, being oriented along one of the Cartesian axes}. We have
shown elsewhere \cite{Nemilentsau2010} that such a boundary-value
problem can be reduced to a Fredholm integral equation of the first
kind for the current density $j^{(\beta)}(z,\mathbf{r}_s,\omega)$.
Thus, we decompose
\begin{equation} \label{Eq:Greent_tensor_SWNT}
\underline{\underline{G}}(\mathbf{r},\mathbf{r}_s,\omega) = \underline{\underline{G}}^{(0)}(\mathbf{r},\mathbf{r}_s,\omega)+
\underline{\underline{G}}^{(sc)}(\mathbf{r},\mathbf{r}_s,\omega),
\end{equation}
where
\begin{equation} \label{Eq:Green_tensor_free}
\underline{\underline{G}}^{(0)}(\mathbf{r},\mathbf{r}_s,\omega) = \left(\underline{\underline{I}} + \frac{\nabla\nabla}{k^2}\right)
\,
\frac{\exp({ik\vert\mathbf{r}-\mathbf{r}_s\vert})}{\vert\mathbf{r}-\mathbf{r}_s\vert}
\end{equation}
is the free-space Green tensor and

\begin{widetext}
\begin{equation}  \label{Eq:Dyson_equation}
G_{\alpha \beta}^{(sc)}(\mathbf{r},\mathbf{r}_s,\omega)= \#e_\alpha\cdot\=G^{(sc)}(\mathbf{r},\mathbf{r}_s,\omega)\cdot\#e_\beta
=\frac{i k^2 R_\mathrm{cn}}{ \omega} \int_{-L/2}^{L/2}j^{(\beta)} (z,\mathbf{r}_s,\omega)
   \int_{0}^{2\pi} G_{\alpha z}^{(0)}(\mathbf{r},\mathbf{r}_{cn},\omega)  d\phi   \, dz\,,
\end{equation}
represents the modification of the free-space Green tensor due to the presence of the SWNT.
The integral equation for $j^{(\beta)}(z,\mathbf{r}_s,\omega)$ is

\begin{equation}
\int_{-L/2}^{L/2} j^{(\beta)}(z',\mathbf{r}_s,\omega)
   {\cal K}(z-z') dz' + C_1 e^{-i k z} + C_2 e^{i k z}     = \frac{1}{2\pi }\int_{-L/2}^{L/2} \frac{e^{i k|z-z'|}}{2 i k}
   \int_{0}^{2\pi} G_{z \beta}^{(0)}(\mathbf{r}'_{cn},\mathbf{r}_s,\omega) d\phi ' dz',
   \label{integral_equation}
\end{equation}
\end{widetext}
where $C_{1}$ and $C_{2}$ are   constants to be determined from the
edge conditions $j^{(\beta)}(\pm
L/2,\mathbf{r}_s,\omega)=0$. The kernel is
\begin{eqnarray}
{\cal K}(z)=\frac{\exp(i k|z|)}{2ik\, \sigma_{zz}(\omega)} - \frac{2
i R_\mathrm{cn}}{\omega} \int_0^{\pi} \frac{e^{i k r}}{r} \,
d\phi\,, \label{Eq:kernel}
\end{eqnarray}
and $r=\sqrt{z^2+4 R_\mathrm{cn}^2 \sin^2(\phi/2)}$.
This integral equation is numerically solvable \cite{Slepyan2006,Nemilentsau2007,Nemilentsau2010}.

\section{Radiation efficiency}
\label{sec:theory_rad}
According to Chap. 8 of Ref.~\onlinecite{Novotny}, in the weak emitter-field coupling regime we can consider an emitter as a classically oscillating at angular frequency $\omega_e$. The radiation efficiency $\eta$ of the emitter in the presence of an SWNT is defined as
\begin{equation} \label{Eq:radiation_efficiency}
\eta = \frac{P_{\mathrm{rad}}}{P_{\mathrm{rad}}+P_{\mathrm{nr}}},
\end{equation}
where $P_{\mathrm{rad}}$ is the power radiated jointly by by the emitter and the SWNT  in the far-field zone and $P_{\mathrm{nr}}$ is the power absorbed by the SWNT. In order to calculate $P_{\mathrm{rad}}$ and $P_{\mathrm{nr}}$, we need to solve the problem of the scattering by the SWNT  of the EMF
radiated by the emitter. The electric field   is the solution of Eq. (\ref{Eq:Green_vector}) with $\mathbf{e}_{\beta}$ on the right side of the equation replaced by $k^2 \mathbf{p}_0$. It also has to satisfy the boundary conditions (\ref{Eq:boundary1}) and (\ref{Eq:boundary2}) on the SWNT surface. This problem has been studied in detail in our previous paper \cite{Nemilentsau2010} and here we present only the main results.

The electric current $\mathbf{J}^{eq}(z) = J^{eq}(z)\mathbf{e}_z $ induced on the SWNT surface by the emitter is the solution of integral equation (\ref{integral_equation}) with $G_{z \beta}^{(0)}(\mathbf{r}'_{cn},\mathbf{r}_s,\omega)$ on the right side of the equation replaced by $k^2 \mathbf{e}_z \cdot \underline{\underline{G}}^{(0)}(\mathbf{r}'_{cn},\mathbf{r}_s,\omega)\cdot \mathbf{p}_0$. After finding $J^{eq}(z)$,
the calculations of $P_{\mathrm{rad}}$ and $P_{\mathrm{nr}}$ are straightforward. Thus,
\begin{equation}
P_{\mathrm{nr}} = \pi R_{cn}\mathrm{Re}\left[\frac{ 1}{\sigma_{zz}} \int\limits_{-L/2}^{L/2} |J^{eq}(z)|^2 dz\right]
\end{equation}
and, with reference to a spherical coordinate system with origin at the centroid of the SWNT and $\#e_{r,\theta,\phi}$ as its unit vector,
\begin{widetext}
\begin{equation}
P_{\mathrm{rad}} = \frac{c r^2}{8\pi} \int\limits_{0}^{\pi} d\theta \sin\theta \int\limits_{0}^{2\pi} d\phi \, \mathbf{e}_{r} \cdot \mathrm{Re}\left\{
\frac{i}{k}\mathbf{E}_{\mathrm{far}}(\mathbf{r})\times \left[ \nabla \times \#E_{\mathrm{far}}(\#r)\right]^\ast\right\} \,,
\end{equation}
where the electric  field in  the far-field zone ($k r \gg 1$) is given by
\begin{equation} \label{Eq21}
\#E_{\mathrm{far}}(\#r)  \simeq  \frac{e^{i k r}}{r}\left[ k^2 e^{-ik\#e_r\cdot\#r_s}
(\#e_\theta\#e_\theta +\#e_\phi\#e_\phi)\cdot\#p_0\,-\#e_{\theta}\frac{i2\pi{R_{cn}}\omega\sin\theta}{c^2}  \,
\int\limits_{-0.5L}^{0.5L} e^{-i k z \cos\theta} J^{eq}(z)dz\,\right] .
 \end{equation}

\end{widetext}

\end{document}